\documentstyle[preprint,aps]{revtex}

\newtheorem{theorem}{Theorem}
\newtheorem{acknowledgement}[theorem]{Acknowledgement}

\begin{document}
\title{Quantum correlations and Nash equilibria of a bi-matrix game}
\author{Azhar Iqbal}
\address{HuMP -- Hull Mathematical Physics\\
Department of Mathematics\\
University of Hull, UK}
\date{May 2004}
\maketitle

\begin{abstract}
Playing a symmetric bi-matrix game is usually physically implemented by
sharing pairs of `objects' between two players. A new setting is proposed
that explicitly shows effects of quantum correlations between the pairs on
the structure of payoff relations and the `solutions' of the game. The
setting allows a re-expression of the game such that the players play the
classical game when{\it \ }their moves are performed on pairs of objects
having correlations that satisfy the Bell's inequalities. If players receive
pairs having quantum correlations the resulting game {\it cannot} be
considered {\it another} classical symmetric bi-matrix game. Also the Nash
equilibria of the game are found to be decided by the nature of the
correlations.
\end{abstract}

\section{Introduction}

Playing a game requires resources for its physical implementation. For
example, to play a bi-matrix game the resources may consist of pairs of
two-valued `objects', like coins, distributed between the players. The
players perform their {\em moves} on the objects and later a referee decides
payoffs after observing them. Game theory usually links players' actions 
{\em directly }to their payoffs, without a reference to the nature of the
objects on which the players have made their moves. Analysis of quantum
games \cite{Eisert} suggests radically different `solutions' can emerge when
the {\em same} game is physically implemented on distributed objects which
are quantum mechanically correlated.

Much of recent work on quantum games \cite{Eisert,Du,Stohler,Li} uses a
particular quantization scheme \cite{Eisert} developed for a bi-matrix game
where two players, on receiving an entangled two-qubit state, play their
moves by local and unitary actions on the state. After disentanglement, a
measurement of the state rewards the players their payoffs. The payoffs
become classical when the moves are performed on a product state. For
example, in Prisoners' Dilemma new and more beneficial equilibrium emerges
in its quantum form \cite{Eisert} when the allowed moves are a {\em chosen}
subset of all possible unitary actions \cite{Benjamin}. Extending the set of
moves to all possible unitary actions results in no equilibrium at all.

Recently Enk and Pike \cite{Enk} have argued that the emergence of new
equilibrium in quantum Prisoners' Dilemma can also be understood as an
equilibrium in a {\em modified} form of the game. They constructed {\em %
another} matrix game, in which players have access to three pure classical
strategies instead of two, claiming that it `captures' everything quantum
Prisoners' Dilemma has to offer. Constructing an extended matrix with an
extra pure move, in their view, is justified because {\em also} in quantum
Prisoners' Dilemma players can play moves which are superpositions of the
two classical moves. Quantum Prisoners' Dilemma, hence, can be thought being
equivalent to playing a modified classical game with an extended matrix.

Truly quantum pairs of objects possess non-local correlations. Though it is
impossible to have a local model, producing {\em exactly} the same data, of
a quantum game set-up but how such unusual correlations may {\em explicitly}
affect solutions of a game when implemented with quantum objects? To how far
extent solutions of a quantum game themselves can be called `truly quantum'
in nature. To address these questions and to find a quantum game for which
it becomes difficult, if not impossible, to construct another classical game
following two {\em constraints} are suggested \cite{QCorrel} that a
quantization scheme should follow:

(C1). In both classical and quantum version of the game the {\em same} set
of moves should be made available to the players.

(C2). The players agree, once and for all, on explicit expressions for their
payoffs which {\em must} not be modified when introducing the quantized
version of the game.

Only the nature of correlations existing between the objects, the players
receive, will now decide whether the resulting game is {\em classical} or 
{\em quantum}. The idea of a `correlation game' \cite{QCorrel}, created to
satisfy the constraints C1 and C2, introduces a new set-up to play bi-matrix
games. Its motivation comes from EPR-type experiments on singlet states
involving correlations of the measurement outcomes. In such experiments the
Bell's inequalities \cite{Peres} are known to be the constraints, derived
under the {\em principle of local causes}, on correlations of measurement
outcomes of two-valued (dichotomic) variables. Truly quantum correlations
are non-local in character and violate the inequalities.

In a quantization scheme, that exploits correlations, players receiving
pairs having local correlations, that do not violate the Bell's inequalties, 
{\em must} result in their payoffs being classical. As pointed out in Ref. 
\cite{QCorrel}, despite explicit dependence of the players' payoffs on
correlations, quantum payoffs can {\em still} be obtained in a correlation
game {\em even} when the correlations do not violate the Bell's
inequalities. In a sense this aspect weakens the argument for a correlation
game. In present paper we try to address this difficulty by following a
different approach in re-defining payoff relations in terms of the
correlations. The new approach is not faced with the indicated difficulty
i.e. local correlations, that do not violate the Bell's inequalities, {\em %
always} result in the classical game. Also in the new approach non-local,
and truly quantum, correlations result in a game that {\em cannot} be
considered just {\em another} classical symmetric bi-matrix game.

\section{Classical symmetric bi-matrix games}

A symmetric bi-matrix game between two players, Alice and Bob, has the
following matrix representation \cite{Rasmusen}

\begin{equation}
\begin{array}{c}
\text{Alice}
\end{array}
\stackrel{
\begin{array}{c}
\text{Bob}
\end{array}
}{
\begin{array}{c}
S_{1} \\ 
S_{2}
\end{array}
\stackrel{
\begin{array}{cc}
S_{1} & S_{2}
\end{array}
}{
\begin{array}{cc}
(r,r) & (s,t) \\ 
(t,s) & (u,u)
\end{array}
}}  \label{matrix}
\end{equation}
where, for example, Alice and Bob get payoffs $s$ and $t$, respectively,
when Alice plays $S_{1}$ and Bob plays $S_{2}$. When Alice and Bob play a
bi-matrix game their moves consist of deciding the probabilities $p$ and $q$%
, respectively, of playing the first strategy $S_{1}$. The second strategy $%
S_{2}$ is then played with probabilities $(1-p)$ and $(1-q)$ respectively.
The mixed strategy payoffs for the players can be written as \cite{QCorrel}

\begin{eqnarray}
P_{A}(p,q) &=&Kpq+Lp+Mq+N  \nonumber \\
P_{B}(p,q) &=&Kpq+Mp+Lq+N  \label{ClassPayoffs}
\end{eqnarray}
where the constants $K,L,M,$ and $N$ can be found in terms of $r,s,t,$ and $%
u $, the coefficients of the bi-matrix. A Nash equilibrium (NE) is a pair $%
(p^{\ast },q^{\ast })$ defined by the following inequalities

\begin{eqnarray}
P_{A}(p^{\ast },q^{\ast })-P_{A}(p,q^{\ast }) &\geq &0  \nonumber \\
P_{B}(p^{\ast },q^{\ast })-P_{B}(p^{\ast },q) &\geq &0  \label{ClassNE}
\end{eqnarray}
For example in the bi-matrix game of Prisoners' Dilemma:

\begin{equation}
\begin{array}{c}
\text{Alice}
\end{array}
\stackrel{
\begin{array}{c}
\text{Bob}
\end{array}
}{
\begin{array}{c}
C \\ 
D
\end{array}
\stackrel{
\begin{array}{cc}
C & D
\end{array}
}{
\begin{array}{cc}
(3,3) & (0,5) \\ 
(5,0) & (1,1)
\end{array}
}}  \label{PDmatrix}
\end{equation}
where $C$ and $D$ represent the strategies of Cooperation and Defection,
respectively, the equilibrium-defining inequalities (\ref{ClassNE}) for the
matrix (\ref{PDmatrix}) are written

\begin{eqnarray}
(p^{\ast }-p)(1+q^{\ast }) &\leq &0  \nonumber \\
(q^{\ast }-q)(1+p^{\ast }) &\leq &0  \label{PDNE}
\end{eqnarray}
giving $p^{\ast }=q^{\ast }=0$ or $(D,D)$ as the unique equilibrium.

\section{Quantum correlation games (QCGs)}

Correlation game \cite{QCorrel} uses an EPR-type setting to play a bi-matrix
game. Repeated measurements are performed on correlated pairs of objects by
two players, each receiving one half. Players Alice and Bob share a
Cartesian coordinate system between them and each player's move consists of
deciding a direction in a given plane. For Alice and Bob these are the $x$-$%
z $ and $y$-$z$ planes respectively.\ Call $\alpha $ and $\beta $ the unit
vectors representing the players' moves. Both players have a choice between
two different orientations i.e. $\alpha $ and $z$ for Alice and $\beta $ and 
$z$ for Bob. Each player measures the angular momentum or spin of his or her
respective half in one of two directions. Let the vectors $\alpha $ and $%
\beta $ make angles $\theta _{A}$ and $\theta _{B}$, respectively, with the $%
z$-axis. To link the players' moves, represented now by angles $\theta _{A}$
and $\theta _{B}$, to the usual probabilities $p$ and $q$ appearing in a
bi-matrix game, an invertible function $g$ is made public at the start of a
game. The $g$-function maps $[0,\pi ]$ to $[0,1]$ and allows to translate
the players' moves to the probabilities $p$ and $q$.

We assume the results of measurements are dichotomic variables, i.e. they
may take only the values $\pm 1$, and are represented by $a,b$ and $c$ for
the directions $\alpha ,\beta $ and the $z$-axis, respectively. Correlations 
$\langle ac\rangle ,\langle cb\rangle $ and $\langle ab\rangle $ can be
found from the measurement outcomes, where the two entries in a bracket
represent the players' chosen directions. When the $z$-axis is shared
between the players as the common direction, Bell's inequality\footnote{%
For perfectly anticorrelated pairs the right hand side of the inequality is $%
1+\left\langle bc\right\rangle $.} is written \cite{Peres} as

\begin{equation}
\left| \left\langle ab\right\rangle -\left\langle ac\right\rangle \right|
\leq 1-\left\langle bc\right\rangle  \label{BellsInequality}
\end{equation}

The classical correlations, written in terms of $\theta _{A}$ and $\theta
_{B}$, are known to be invertible. This fact allows us to express $\theta
_{A}$ and $\theta _{B}$ in terms of the correlations $\langle ac\rangle $
and $\langle cb\rangle $. The $g$-function then makes possible to translate $%
\theta _{A}$ and $\theta _{B}$ to $p$ and $q$, respectively. In effect the
classical bi-matrix payoffs are re-expressed in terms of the classical
correlations $\langle ac\rangle $ and $\langle cb\rangle $. We claim now
that our classical game is given, {\em by definition}, in terms of the
correlations. Such re-expression opens the way to find `quantum' payoffs
when the correlations become quantum mechanical.

In this setting the players' payoffs involve only the correlations $\langle
ac\rangle $ and $\langle cb\rangle $, instead of the three correlations $%
\langle ac\rangle ,\langle cb\rangle $ and $\langle ab\rangle $ present in
the inequality (\ref{BellsInequality}), when $z$-axis is the common
direction between the players. This aspect results in obtaining `quantum'
payoffs even when the correlations are local and satisfy the inequality (\ref
{BellsInequality}). The motivation for introducing EPR-type setting to
bi-matrix games is to exploit quantum correlations to generate quantum
payoffs. So that, when the correlations are local, the classical game must
be produced. We show below the possibility of such a connection by some
modifications in the setting of a correlation game suggested previously. In
the modified setting the classical payoffs are {\em always} obtained
whenever the correlations $\langle ac\rangle ,\langle cb\rangle $ and $%
\langle ab\rangle $ satisfy the Bell's inequality (\ref{BellsInequality}).

\section{A new approach towards defining a correlation game}

Following modifications are suggested in the setting of a correlation game:

\begin{enumerate}
\item  A player's move consists of defining a direction in space by
orientating a unit vector. However, this direction is not confined to only
the $x$-$z$ or $y$-$z$ planes. A player's choice of a direction can be {\em %
anywhere} in three-dimensional space. Therefore, Alice's move is to define a
unit vector $\alpha $ and, similarly, Bob's move is to define a unit vector $%
\beta $.

\item  The $z$-axis is shared between the players as the common direction.

\item  On receiving a half of a correlated pair, a player measures its spin
in two directions. For Alice these directions are $\alpha $ and $z$ and for
Bob these directions are $\beta $ and $z$.

\item  Each player measures spin with {\em equal} probability in his/her two
directions.

\item  Players agree together on {\em explicit expressions} giving their
payoffs $P_{A}$ and $P_{B}$ in terms of all three correlations i.e.
\end{enumerate}

\begin{eqnarray}
P_{A} &=&P_{A}(\langle ac\rangle ,\langle cb\rangle ,\langle ab\rangle ) 
\nonumber \\
P_{B} &=&P_{B}(\langle ac\rangle ,\langle cb\rangle ,\langle ab\rangle )
\label{CorrelPayoffs}
\end{eqnarray}
These modifications eliminate the need for introducing the $g$- functions as
done in Ref. \cite{QCorrel}. The modifications are also consistent with the
constraints C1 and C2 and the idea of a correlation game, developed in Ref. 
\cite{QCorrel}, essentially retains its spirit. More importantly, a player's
move can be {\em any} direction in space.

\subsection{Defining correlation payoffs in the new approach}

A possible way is shown now to define the correlation payoffs (\ref
{CorrelPayoffs}) which reduce to the classical payoffs (\ref{ClassPayoffs})
whenever the correlations $\langle ab\rangle ,\langle ac\rangle $ and $%
\langle bc\rangle $ satisfy the inequality (\ref{BellsInequality}).

Consider two quantities $\varepsilon $ and $\sigma $ defined as follows

\begin{equation}
\varepsilon =\sqrt{3+\langle bc\rangle ^{2}+2\langle ab\rangle \langle
ac\rangle },\text{ \ \ \ }\sigma =\sqrt{2(1+\langle bc\rangle )+\langle
ab\rangle ^{2}+\langle ac\rangle ^{2}}  \label{A&B}
\end{equation}
The quantities $\varepsilon $ and $\sigma $ can adapt only real values
because the correlations $\langle ac\rangle ,\langle cb\rangle ,$ and $%
\langle ab\rangle $ are always in the interval $\left[ -1,1\right] $.
Consider now the quantities $(\varepsilon -\sigma )$ and $(\varepsilon
+\sigma )$. By definition $\varepsilon $ and $\sigma $ are non-negative,
therefore, the quantity $(\varepsilon +\sigma )$ always remains
non-negative. It is observed that if $0\leq $ $(\varepsilon -\sigma )$ then
the correlations $\langle ac\rangle ,\langle cb\rangle $ and $\langle
ab\rangle $ satisfy the inequality (\ref{BellsInequality}). It is because if 
$0\leq (\varepsilon -\sigma )$ then $0\leq (\varepsilon +\sigma
)(\varepsilon -\sigma )=\varepsilon ^{2}-\sigma ^{2}$. But $\varepsilon
^{2}-\sigma ^{2}=(1-\langle bc\rangle )^{2}-\left| \langle ab\rangle
-\langle ac\rangle \right| ^{2}$ so that $\left| \langle ab\rangle -\langle
ac\rangle \right| ^{2}\leq (1-\langle bc\rangle )^{2}$ which results in the
inequality (\ref{BellsInequality}). All the steps in the proof can be
reversed and it follows that when the correlations $\langle ac\rangle
,\langle cb\rangle $ and $\langle ab\rangle $ satisfy the Bell's inequality,
the quantity $(\varepsilon -\sigma )$ remains non-negative

For a singlet state satisfying the inequality (\ref{BellsInequality}) both
the quantities $(\varepsilon +\sigma )$ and $(\varepsilon -\sigma )$ are
non-negative and must have maxima. Hence, it is possible to find two
non-negative numbers $\frac{(\varepsilon -\sigma )}{\max (\varepsilon
-\sigma )}$ and $\frac{(\varepsilon +\sigma )}{\max (\varepsilon +\sigma )}$
in the range $\left[ 0,1\right] $, whenever the inequality (\ref
{BellsInequality}) holds. Because $0\leq \varepsilon ,\sigma \leq \sqrt{6}$
we have $\max (\varepsilon -\sigma )=\sqrt{6}$ and $\max (\varepsilon
+\sigma )=2\sqrt{6}$. The numbers $(\varepsilon -\sigma )/\sqrt{6}$ and $%
(\varepsilon +\sigma )/2\sqrt{6}$ are in the range $[0,1]$ when the
inequality holds. These numbers are also {\em independent} from each other.

The above argument paves the way to associate a pair $(p,q)$ of independent
numbers to the players' moves $(\alpha ,\beta )$, that is

\begin{equation}
p=p(\alpha ,\beta )\text{, \ \ \ \ \ \ }q=q(\alpha ,\beta )
\label{ProbDirLink}
\end{equation}
where $p,q$ are in the interval $[0,1]$ for all directions $\alpha ,\beta $,
when the input states do not violate the inequality (\ref{BellsInequality}).
From the pair $(p,q)$ a directional pair can also be found as

\begin{equation}
\alpha =\alpha (p,q)\text{, \ \ \ \ \ \ }\beta =\beta (p,q)
\label{ProbDirLink1}
\end{equation}
but more than one pair $(\alpha ,\beta )$ of directions may correspond to a
given pair of numbers. The converse, however, is not true for known input
states. That is, for known input states, only one pair $(p,q)$ can be
obtained from a given pair $(\alpha ,\beta )$ of directions.

Players' payoffs can now be expressed in a {\em correlation form} by the
following replacements

\begin{equation}
p(\alpha ,\beta )\sim (\varepsilon -\sigma )/\sqrt{6},\text{ \ \ }q(\alpha
,\beta )\sim (\varepsilon +\sigma )/2\sqrt{6}  \label{Replacements}
\end{equation}
leading to a re-expression of the classical payoffs (\ref{ClassPayoffs}) as

\begin{eqnarray}
P_{A}(\alpha ,\beta ) &=&Kp(\alpha ,\beta )q(\alpha ,\beta )+Lp(\alpha
,\beta )+Mq(\alpha ,\beta )+N  \nonumber \\
P_{B}(\alpha ,\beta ) &=&Kp(\alpha ,\beta )q(\alpha ,\beta )+Mp(\alpha
,\beta )+Lq(\alpha ,\beta )+N
\end{eqnarray}
or more explicitly as

\begin{eqnarray}
P_{A}(\alpha ,\beta ) &=&\frac{K}{12}(\varepsilon ^{2}-\sigma ^{2})+\frac{L}{%
\sqrt{6}}(\varepsilon -\sigma )+\frac{M}{2\sqrt{6}}(\varepsilon +\sigma )+N 
\nonumber \\
P_{B}(\alpha ,\beta ) &=&\frac{K}{12}(\varepsilon ^{2}-\sigma ^{2})+\frac{M}{%
\sqrt{6}}(\varepsilon -\sigma )+\frac{L}{2\sqrt{6}}(\varepsilon +\sigma )+N
\label{Qpayoffs}
\end{eqnarray}
where a player's payoff now depends on the direction s/he has chosen. The
payoffs (\ref{Qpayoffs}) are obtained under the constraints C1 and C2 and
are now functions of all the three correlations.

The relations (\ref{ProbDirLink}) can also be imagined as follows. When
Alice decides a direction $\alpha $ in space, it corresponds to a curve in
the $p$-$q$ plane. Similarly, Bob's decision of the direction $\beta $
defines another curve in the $p$-$q$ plane. The relations (\ref{Replacements}%
) assure that only one pair $(p,q)$ can then be obtained as the intersection
between the two curves.

The set-up assures that for input states satisfying the inequality (\ref
{BellsInequality}), all of the players' moves $(\alpha ,\beta )$ result in
the correlation payoffs (\ref{Qpayoffs}) generating identical to the
classical payoffs (\ref{ClassPayoffs}). For such input states the relations (%
\ref{Replacements}) give the numbers $p,q$ in the interval $[0,1]$, which
can then be interpreted as probabilities. However, for input states
violating the inequality (\ref{BellsInequality}), a pair $(p,q)$ $\in
\lbrack 0,1]$ {\em cannot} be associated with players' moves $(\alpha ,\beta
)$. It is because for such states the quantity $(\varepsilon -\sigma )$
becomes negative and the correlation payoffs (\ref{Qpayoffs}) generate
results having a different form from the classical payoffs (\ref
{ClassPayoffs}).

\section{Nash equilibria of QCGs}

Because the players' moves consist of defining directions in space, the Nash
inequalities can be written as

\begin{eqnarray}
P_{A}(\alpha _{0},\beta _{0})-P_{A}(\alpha ,\beta _{0}) &\geq &0  \nonumber
\\
P_{B}(\alpha _{0},\beta _{0})-P_{B}(\alpha _{0},\beta ) &\geq &0
\label{NEdirections}
\end{eqnarray}
where the pair $(\alpha _{0},\beta _{0})$ corresponds to the pair $(p^{\ast
},q^{\ast })$, defined in (\ref{ClassNE}), via the relations (\ref
{Replacements}). The inequalities (\ref{NEdirections}) are same as the
inequalities (\ref{ClassNE}), except their re-expression in terms of the
directions.

When the correlations in the input state violate the inequality (\ref
{BellsInequality}), the payoff relations (\ref{Qpayoffs}) also lead to
disappearance of the classical equilibria. It can be seen, for example, by
considering the Nash inequalities for the Prisoners' Dilemma (\ref{PDNE}).
Let the directional pair $(\alpha _{D},\beta _{D})$ correspond to the
equilibrium $(D,D)$, that is, the inequalities (\ref{NEdirections}) are
written as

\begin{eqnarray}
P_{A}(\alpha _{D},\beta _{D})-P_{A}(\alpha ,\beta _{D}) &\geq &0  \nonumber
\\
P_{B}(\alpha _{D},\beta _{D})-P_{B}(\alpha _{D},\beta ) &\geq &0
\label{NEDD}
\end{eqnarray}
Assume the players receive input states that violate the inequality (\ref
{BellsInequality}). It makes the quantity $(\varepsilon -\sigma )<0$, that
is, the players' moves $\alpha $ and $\beta $ will not correspond to a point
in the $p$-$q$ plane where $p,q\in \lbrack 0,1]$. Also the directional pair $%
(\alpha _{D},\beta _{D})$ does not remain a NE. It is because the pair{\it \ 
}$(\alpha _{D},\beta _{D})${\it \ }is a NE {\em only} if players' choices of 
{\em any} directional pair{\it \ }$(\alpha ,\beta )${\it \ }corresponds to a
point in the{\it \ }$p${\it -}$q${\it \ }plane where{\it \ }$p,q\in \lbrack
0,1]$. Because for input states that violate the inequality (\ref
{BellsInequality}) a pair of players' moves $(\alpha ,\beta )$ does not
correspond to a point in the $p$-$q$ plane with $p,q\in \lbrack 0,1]$,
hence, the directional pair $(\alpha _{D},\beta _{D})$ does not remain a NE
in the quantum game. The disappearance of the classical equilibrium now
becomes {\it linked} with the violation of the inequality (\ref
{BellsInequality}) by the correlations in the input states.

\section{Quantum game as another classical game?}

Coming back to the questions raised in the Introduction, we now try to
construct a classical bi-matrix game, corresponding to a quantum game,
resulting from the payoff relations (\ref{Qpayoffs}). The classical game is
assumed to have the {\em same} general structure of players' payoffs as
given in Eqs. (\ref{ClassPayoffs}). This assumption derives from the hope
that the quantum game, corresponding to correlations in the input states
that violate the inequality (\ref{BellsInequality}), is also equivalent to 
{\em another} symmetric bi-matrix game. It is shown below that such a
construction cannot be permitted.

Suppose the input states violate the inequality (\ref{BellsInequality}). For 
{\em any} direction Alice chooses to play, her payoff given by Eqs. (\ref
{Qpayoffs}) can also be written as

\begin{equation}
P_{A}(\alpha ,\beta )=K^{\prime }pq+L^{\prime }p+Mq+N
\label{Alice'sNewPayoff}
\end{equation}
where $K^{\prime }=-K$ and $L^{\prime }=-L$ and $p,q\in \lbrack 0,1]$.
Assuming that the constants $K^{\prime },L^{\prime },M,$ and $N$ define a
`new' symmetric bi-matrix game the Bob's payoff should then be written as

\begin{equation}
P_{B}(p,q)=K^{\prime }pq+Mp+L^{\prime }q+N  \label{Bob'sNewPayoff1}
\end{equation}
But in fact (\ref{Bob'sNewPayoff1}) is not obtained as the Bob's payoff in
the quantum game with correlations violationg the inequality (\ref
{BellsInequality}). Bob's payoff in the quantum game is given as

\begin{equation}
P_{B}(p,q)=K^{\prime }pq+M^{\prime }p+Lq+N  \label{Bob'sNewPayoff2}
\end{equation}
where $M^{\prime }=-M$. Hence the game resulting from the presence of
quantum correlations in the input states {\em cannot} simply be explained as
another classical symmetric bi-matrix game: a game obtained by defining new
coefficients of the matrix involved. Players' payoffs in the quantum game
reside outside the structure of payoffs of a classical symmetric bi-matrix
game. The payoffs can be explained within this structure {\em only} by
invoking negative probabilities.

An asymmetric bi-matrix game can, of course, be constructed having identical
solutions to the quantum game. In fact for {\em any} quantum game a
classical model can {\em always} be constructed that summarizes the complete
situation and has identical to the quantum solutions as far as the players'
payoffs are concerned. A model that relates players' moves directly to their
payoffs in accordance with the usual approach in game theory. But still it
is not an answer to our initial question: how solutions of a game are
affected by the presence of quantum correlations between the physical
objects used to implement the game. It is because the question can then
simply be rephrased as: what if the modified classical game is played with
physical objects having quantum correlations.

\section{Summary}

The idea of a correlation game is about re-expression of payoffs of a
classical bi-matrix game in terms of correlations of measurement outcomes
made on pairs of correlated particles. The measurement outcomes are
dichotomic variables and their correlations are obtained by averaging over a
large number of pairs. Bell's inequalities represent constraints on these
correlations obtained under the principle of local causes. A re-expression
of the classical payoffs of a bi-matrix game in terms of correlations opens
the way to explicitly see the effects of quantum correlations on the
solutions of the game.

In this paper a new setting is proposed where two players play a bi-matrix
game by repeatedly performing measurements on correlated pairs of objects.
The setting is motivated by EPR-type experiments performed on singlet
states. On receiving a half of a pair, a player makes a measurement of its
spin in one of the two directions available to him or her. The measurements
are performed with {\em equal probability} in the two directions. Both
players share a common direction and defining the {\em other} direction is a
player's {\em move}.

We show how within this set-up a correlation version of a symmetric
bi-matrix game can be defined. The correlation game shows some interesting
properties. For example, it reduces to the corresponding classical game when
the correlations in the input states are local and do not violate the Bell's
inequality (\ref{BellsInequality}). However, when the inequality is
violated, the stronger correlations generate results that can be understood,
within the structure of classical payoffs in a symmetric bi-matrix game, 
{\em only} by invoking negative probabilities. It is shown that a classical
Nash equilibrium is affected when the game is played with input states
having quantum correlations. The proposed set-up also provides a new
perspective on the possibility of reformulating the Bell's inequalities in
terms of a bi-matrix game played between two spatially-separated players.

\begin{acknowledgement}
Author gratefully acknowledges motivating and helpful discussions with Dr.
Stefan Weigert.
\end{acknowledgement}

\end{document}